%% file: main.tex
\documentclass[conference,compsoc]{IEEEtran}

\ifCLASSOPTIONcompsoc
  \usepackage[nocompress]{cite}
\else
  \usepackage{cite}
\fi

\ifCLASSINFOpdf
\else
\fi

\ifCLASSOPTIONcompsoc
  \usepackage[caption=false,font=footnotesize,labelfont=sf,textfont=sf]{subfig}
\else
  \usepackage[caption=false,font=footnotesize]{subfig}
\fi

\usepackage{soul}

\usepackage{tikz}
\newcommand{\circlednum}[1]{\protect\circlednumAux{#1}}

\newcommand{\circlednumAux}[1]{%
  \tikz[baseline=(char.base)]%
    \node[shape=circle, draw=black, fill=black, inner sep=0.6pt, scale=0.85] (char) {\textcolor{white}{#1}};%
}
\usetikzlibrary{shapes.geometric, arrows, positioning}

\tikzstyle{startstop} = [rectangle, rounded corners, minimum width=2.8cm, minimum height=1cm, text centered, draw=black, fill=gray!20]
\tikzstyle{process} = [rectangle, minimum width=2.8cm, minimum height=1cm, text centered, draw=black, fill=blue!15]
\tikzstyle{decision} = [diamond, minimum width=3.2cm, minimum height=1cm, text centered, draw=black, fill=blue!15]
\tikzstyle{input} = [rectangle, minimum width=1.8cm, minimum height=0.8cm, text centered, draw=black, fill=white]
\tikzstyle{arrow} = [thick,->,>=stealth]

\newcommand{\taggedpara}[1]{\noindent\textbf{#1.}}

\usepackage{xcolor}

\usepackage{fancyhdr}

\fancypagestyle{firstpage}{
  \fancyfoot[L]{\textcolor{gray}{\small Author's version, to appear, 2026 IEEE Symposium on Security and Privacy (SP)}}
  \fancyfoot[C]{}
  \fancyfoot[R]{}
  
}

\usepackage{listings}
\usepackage{hyperref}
\begin{document}

\title{AEX-NStep: Probabilistic Interrupt Counting Attacks on Intel SGX}

\author{
\IEEEauthorblockN{
Nicolas Dutly, Friederike Groschupp, Ivan Puddu, Kari Kostiainen, Srdjan Čapkun
}
\IEEEauthorblockA{
Department of Computer Science, ETH Zurich \\
\{nicolas.dutly, friederike.groschupp, ivan.puddu, kari.kostiainen, srdjan.capkun\}@inf.ethz.ch
}
}

\maketitle

\begin{abstract}
\input{content/00-abstract}
\end{abstract}

\thispagestyle{firstpage}

\IEEEpeerreviewmaketitle

\section{Introduction}
\input{content/01-introduction}

\section{Background}\label{background}

\input{content/02-background}

\section{Probabilistic Interrupt Counting}\label{probcountingattacks}

\taggedpara{Problem Statement} 

In this work, we ask the following questions: Is AEX-Notify, and the prevention of deterministic single stepping, sufficient to prevent practical interrupt counting attacks on SGX enclaves? Can we improve the analysis of interrupt counting attacks beyond the approaches used to evaluate AEX-Notify~\cite{aex-notify}? \\

\begin{figure}
    \centering
    \subfloat[Intuition of SGX-step and its mitigation AEX-Notify]{   
    \begin{minipage}{0.45\textwidth}
        \centering
        \includegraphics[width=\linewidth]{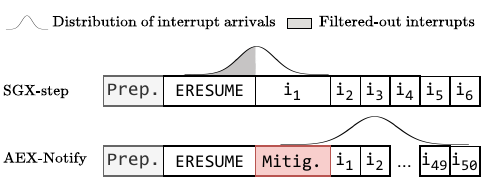}
        \label{fig:intuition-before}
    \end{minipage}}
    
      \subfloat[Intuition of our attacks on AEX-Notify-enabled enclaves]{
     \begin{minipage}[b]{0.45\textwidth}
        \centering
        \includegraphics[width=\linewidth]{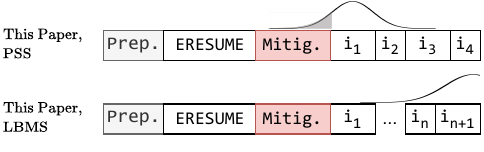}
        \label{fig:intuition-us}
    \end{minipage}}   
    
    \caption{Comparison of our attack primitives to SGX-Step and AEX-Notify. The attacker resumes the enclave with ERESUME after finishing with the necessary preparations (Prep.). (a) SGX-Step: The attacker extends the execution time of the first enclave instruction, arms a timer-based interrupt, and filters out zero-steps by examining PTE attributes. AEX-Notify introduces a mitigation phase (Mitig.), which ensures that instructions are pre-fetched, reducing their execution time and preventing enclave progress from being determined through page table entries (PTEs). (b) We extend the execution time of enclave instructions by disabling caches and use IPIs to reduce the variability of interrupt arrivals. For PSS, we align the arrival distribution such that interrupts land in the mitigation or at the first enclave instructions. Interrupts hitting the mitigation are filtered out by breaking obfuscated forward progress. For LBMS, we align the interrupt arrival redistribution such that interrupts trigger after at least $n$ instructions.}
   \label{fig:comp-sgx-aex}
\end{figure}

\taggedpara{SGX-Step and AEX-Notify}
In Fig.~\ref{fig:comp-sgx-aex}, we illustrate the main differences between SGX (step), AEX-Notify-enabled SGX, and AEX-NStep. 
SGX-Step extends the window of the  enclave instruction by forcing the processor to prefetch its working set, making it possible for the instruction to be interrupted reliably. It further filters out zero-steps by examining page table attributes.

AEX-Notify addresses SGX-Step attacks by inserting a prefetching mitigation stub before resuming an enclave, narrowing the instruction execution window, and preventing interrupts from precisely landing during the execution of the next enclave instruction. Interrupts now potentially skip many instructions and can also occur during mitigation. AEX-Notify's obfuscated forward progress guarantee attempts to prevent attackers from filtering interrupts landing during the mitigation. The security analysis of AEX-Notify~\cite{aex-notify} concludes that these changes break conventional deterministic single-stepping.

\begin{figure*}[t]
    \centering
    \includegraphics[width=0.94\linewidth]{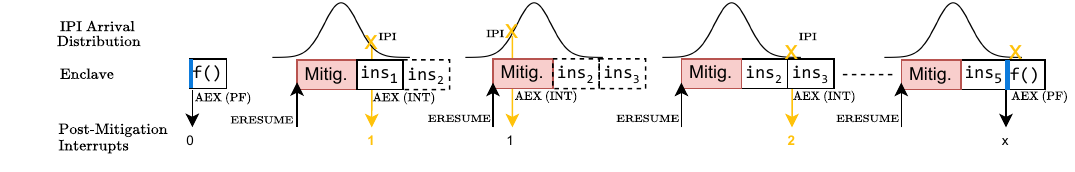}
    \caption{Conceptual overview of probabilistic single-stepping. The attacker wants to determine which branch is taken in Listing~\ref{list:secret_pss}. The vulnerable branch is delimited by accesses to another page (\texttt{f()}), which allows the attacker to locate the vulnerable code section with the help of page faults. Enclave resumption after an AEX will lead to the execution of the AEX-Notify mitigation (Mitig.). When an AEX is encountered, the attacker resumes the enclave and fires an IPI from a concurrent thread, which is calibrated such that the right tail of the interrupt's arrival distribution overlaps with the first instructions executed after the AEX-Notify mitigation. The attacker counts the interrupts observed in the vulnerable code section for each bit of the secret. On average, non-zero secret bits entail more interrupts, as the corresponding branch executes more instructions. }
    \label{fig:concept-pss}
\end{figure*}

\taggedpara{Intuition of AEX-NStep}
AEX-NStep extends the attacker's capabilities to counteract the AEX-Notify mitigation (Fig.~\ref{fig:intuition-us}). We extend the execution window of enclave instructions by disabling caches on the enclave core, use IPIs fired from a concurrent thread, and break AEX-Notify's obfuscated forward progress guarantee. We discuss these techniques in detail in Section~\ref{sec:details}. They play a key role in enabling attackers to influence the number of instructions executed between interrupts, as well as providing fine-grained control over interrupt arrival distributions. 
In our attacks, we adjust the interrupt arrival distribution to achieve various objectives. Shifting the distribution to the left ensures that if an interrupt arrives after the mitigation, it will arrive within a couple of enclave instructions. Combined with our ability to detect and filter interrupts that land during mitigation, this process effectively approximates the behavior of deterministic single-stepping. We name this probabilistic single-stepping (PSS). Shifting the IPI arrival distribution to the right enables us to observe interrupts only when specific branches are taken, a primitive that we refer to as lower-bounded multi-stepping (LBMS). 

PSS is well-suited for targeting secrets that remain constant between runs, while LBMS can target ephemeral secrets that change with every run.
In this section, we illustrate both primitives on an example in the context of an AEX-Notify-enabled SGX enclave.

\taggedpara{Threat Model} We adopt the same threat model under which AEX-Notify~\cite{aex-notify} was developed: the attacker has complete and privileged control over the operating system and non-enclave software. An attacker may enable or disable CPU cores, choose which cores to schedule enclave threads on, enable or disable features such as caching and hyperthreading (SMT), and, in general, tamper with any software not running in an enclave. We further assume that an attacker can deterministically single-step any enclave threads for which AEX-Notify is not enabled~\cite{DBLP:conf/sosp/BulckPS17}. Furthermore, the attacker may infer page-level progress of SGX threads to determine when to start and stop counting instructions by tracking cross-page memory accesses, as is commonly assumed by existing research~\cite{frontal, tdx-down}. The CPU package and microcode are considered to be non-compromised. The attacker has prior knowledge of the vulnerable code or a copy of the victim enclave, which aligns with the SGX threat model.

\subsection{Probabilistic Single-Stepping (PSS)}
We assume that the instruction windows have been extended and that the attacker can filter out any interrupts that occur during the mitigation. We discuss how we implement these abilities on AEX-Notify-enabled enclaves in Section~\ref{sec:details-pss}.
We illustrate the principle of PSS by targeting a secret-dependent branch, such as the one in Listing~\ref{list:secret_pss}. The victim executes five instructions for non-zero secret bits and two otherwise. Such issues are not uncommon, as compilers can unintentionally introduce branches or extra instructions despite developers' efforts to balance code~\cite{DBLP:journals/corr/abs-2410-13489, DBLP:conf/uss/MoghimiBHPS20, frontal}.

\taggedpara{Requirements} The vulnerable branch must depend on a static secret that does not vary between executions, as probabilistic single-stepping relies on multi-trace averaging to extract secrets. While multi-trace averaging is also used in the context of conventional timing attacks, we target branches that differ by potentially only a few instructions, which falls below the noise threshold in conventional timing attacks. This especially holds when considering the unpredictable latency of instructions such as \texttt{ERESUME}, which can take up to one thousand CPU cycles to execute~\cite{aex-notify}.

\taggedpara{Offline Phase}
Probabilistic single-stepping aims to approximate deterministic single-stepping by landing interrupts as close as possible to the EARP. As noted by~\cite{aex-notify}, interrupts are typically normally distributed with a variance of up to a couple of hundred cycles. Before targeting the victim, the attacker must calibrate interrupts such that the right tail of the interrupt arrival distribution aligns with EARP. Calibrating the IPI arrival distribution can be achieved by running an attacker-controlled debug enclave and inspecting where interrupts land. Given that the AEX-Notify mitigation is largely independent of the target enclave, this calibration is transferable to different targets.

\taggedpara{Attack Flow}
\begin{figure}
\renewcommand{\figurename}{Listing}
\begin{lstlisting}[language=Python]
for bit in secret:
  f() # NX, different code page
  if (bit):
    ins_1
    ins_2
    ins_3
    ins_4
    ins_5
  else
    ins_a
    ins_b
  f() #  NX, different code page
\end{lstlisting}
\caption{Secret-dependent branch with a static secret. Attackers can use cross-page memory accesses to isolate the vulnerable branch, represented here by function calls residing on different pages.
}
\label{list:secret_pss}
\end{figure}
Following the visualization provided in Fig.~\ref{fig:concept-pss}, the attacker begins by advancing the enclave to the start of the vulnerable branch. After resuming the enclave, the attacker fires an IPI after a pre-calibrated fixed delay from a concurrent thread. Following the example in Fig.~\ref{fig:concept-pss}, the first interrupt lands during the execution of \texttt{ins\_1}, triggering an AEX.\footnote{For this presentation, we assume that the attacker can break AEX-Notify's obfuscated forward progress guarantee, and thus detect when the enclave makes progress. In Section~\ref{sec:details-pss} we show how to do so in practice.}
The attacker increments an interrupt counter variable, resumes the enclave's control flow, and repeats the fixed-delay IPI procedure from the concurrent thread. The next interrupt lands during the mitigation, causing an AEX. The attacker does not count the interrupt, as it landed during the mitigation. This process is repeated until the enclave attempts to execute the second \texttt{f()} call, which is detected using conventional page-fault side-channels. The attacker saves the current post-mitigation interrupt count, resets the counter, and repeats the process for every subsequent secret bit, counting the number of post-mitigation interrupts that occur. Since the secret is static, the attacker can repeat this process, collecting multiple interrupt counts for every bit. Since non-zero secret bits entail executing more instructions, the attacker will, on average, observe more post-mitigation interrupts for them. By averaging observed interrupt counts, the attacker can recover the entire secret.

\subsection{Lower-Bounded Multi-Stepping (LBMS)}
\label{section:lbms_concept}
\begin{figure}
    \renewcommand{\figurename}{Listing}
    \centering
\begin{lstlisting}[language=Python]
def ecdsa_sign(msg):
    k = rnd_nonce()
    f()  # NX, different code page
    if MSB(k) == 0xFFFF:
        ins_1
        ins_2
        ins_3
        ins_4
        ins_5
    else:
        ins_a
        ins_b
    f()  # NX, different code page
    return signature
\end{lstlisting}
    \caption{Secret-dependent branch with an ephemeral secret. The branch executes different numbers of instructions depending on the most significant bits of a nonce $k$. If the nonce is biased (i.e, its most significant bits are \texttt{0xFFFF}), more instructions will be executed between the two page fault triggers than in the non-biased case. }
   \label{secret_lbms}
\end{figure}

\begin{figure*}
    \centering
    \subfloat[Signed with unbiased nonce: No interrupt observed]{   
    \begin{minipage}[l]{0.45\textwidth}
        \centering
        \includegraphics[width=\linewidth]{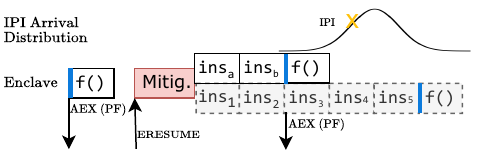}
        \label{fig:lbms-unbiased}
    \end{minipage}}%
      \subfloat[Signed with biased nonce: Interrupt observed before page fault]{
     \begin{minipage}[r]{0.45\textwidth}
        \centering
        \includegraphics[width=\linewidth]{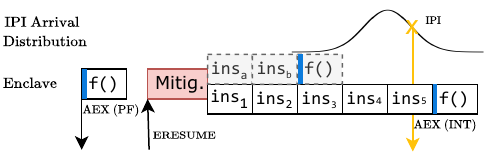}
        \label{fig:lbms-biased}
    \end{minipage}}   
    
    \caption{Conceptual overview of an attack leveraging LBMS to detect ECDSA signatures signed with biased nonces. In contrast to PSS, LBMS takes a one-shot approach which does not require averaging: By lower-bounding the interrupt arrival time to arrival at least after two full instructions have been executed, an interrupt will only be observed before reaching the second boundary function call if the longer branch was taken, which in this case corresponds to the usage of a biased nonce during the signing process. Not observing an interrupt does not imply that the nonce is biased, as the interrupt arrival distribution does not fit perfectly within the extra instructions. As such, interrupts may land after the second boundary function call.}
   \label{list:concept-lbms}
\end{figure*}

Probabilistic Single-Stepping is not well-suited for targeting ephemeral secrets. We introduce Lower-Bounded Multi-Stepping (LBMS), an interrupt counting primitive that does not rely on averaging interrupt counts over multiple traces. In practice, some targets may depend on ephemeral secrets that change with every call of the victim. Listing~\ref{secret_lbms} illustrates such a case, where extra instructions are executed if a nonce used to generate an ECDSA signature is biased. The attacker may want to detect the execution of these additional instructions, as collecting a sufficient number of biased signatures can result in a full private key recovery.

\taggedpara{Requirements} LBMS does not require a fixed secret. However, we cannot target all unbalanced branches, as we can only detect the execution of the longest branch. LBMS cannot be used to target cases where the shorter branch leaks information. For example, in Listing~\ref{secret_lbms}, the longer branch corresponds to the case where the nonce used to sign the current signature is biased. An attacker can exploit this leak to reconstruct private signing keys. As such, Listing~\ref{secret_lbms} can be exploited using LBMS. To correctly time interrupts, we require a target-dependent calibration step. 

\taggedpara{Offline Phase}
In PSS the IPI calibration is target-agnostic. In LBMS, the IPI calibration must be adapted for each target. The interrupt arrival distribution must be calibrated such that interrupts arrive only after at least $n$ post-mitigation instructions have been executed, where $n$ is the number of instructions in the shorter, secret-dependent branch. Following the example in Listing~\ref{secret_lbms}, the calibration ensures that at least two instructions are executed before an interrupt occurs. This calibration can be performed on an attacker-controlled debug enclave containing a copy of the relevant victim instructions. We assume the first $n$ instructions of both branches have similar cycle requirements. We further discuss the impact of this assumption in Section~\ref{limitations}.

\taggedpara{Attack Flow}
 We follow the attack flow depicted in Fig.~\ref{list:concept-lbms}. The LBMS attack begins with the same page-fault-based operations used in PSS to identify when the target branch is executed. Once this process is completed, the attacker will resume the enclave and fire a fixed-delay interrupt from a concurrent thread. The delay is calibrated such that at least two instructions are guaranteed to be executed after the mitigation has taken effect. Following the example in Fig.~\ref{list:concept-lbms}, the attacker will either observe the second \texttt{f()} call or an AEX caused by the previously fired IPI. If the interrupt causes the AEX, then the enclave must have executed the longer conditional branch, which implies that the nonce used for the current ECDSA signature is biased.
 
\subsection{Towards Practical Attacks}

The attack intuitions provided in this section abstract away several non-trivial, technical details, such as how attackers can extend instruction execution windows despite the AEX-Notify prefetching phase and how attackers can detect and filter interrupts that land during the mitigation to break AEX-Notify's obfuscated forward progress guarantee. We present solutions to these challenges in Section~\ref{sec:details}.

To understand the practical impact of these attack primitives, we first evaluate the expected success rates for different generic branch properties (see Section~\ref{eval}) and then demonstrate concrete attacks by replicating SGX-Step's \texttt{memcmp} attack~\cite{vanbulck_sgxstep_memcmp} using PSS and leaking ECDSA private keys using LBMS  (see Section~\ref{leakage}). These attacks demonstrate that AEX-Notify is insufficient to prevent all interrupt counting attacks, suggesting its threat model may be too broad to protect SGX enclaves effectively.

\input{content/03-attack-overview}

\section{Evaluating Probabilistic Stepping Rates}\label{eval}

\input{content/04-attack-analysis}

\section{Practical Attacks}\label{leakage}

\input{content/05-exploitation}

\section{Discussion}

\input{content/06-limitations}\label{limitations}

\section{Related Work}

\input{content/07-related-work}

\section{Conclusion}
\label{conclusion}
\input{content/08-conclusion}

\section{Acknowledgments}

This work was partially supported by the Zurich Information Security and Privacy Center (ZISC).

\bibliographystyle{plain}
\bibliography{bibliography}

\end{document}

%% file: content/00-abstract.tex
To mitigate interrupt-based stepping attacks (notably using SGX-Step), Intel introduced AEX-Notify, an ISA extension to Intel SGX that aims to prevent deterministic single-stepping.
In this work, we introduce AEX-NStep, the first interrupt counting attack on AEX-Notify-enabled Enclaves. We show that deterministic single-stepping is not required for interrupt counting attacks to be practical and that, therefore, AEX-Notify does not entirely prevent such attacks.
We specifically show that one of AEX-Notify's security guarantees, obfuscated forward progress, does not hold, and we introduce two new probabilistic interrupt counting attacks. We use these attacks to construct a practical ECDSA key leakage attack on an AEX-Notify-enabled SGX enclave.
Our results extend the original security analysis of AEX-Notify and inform the design of future mitigations.

%% file: content/01-introduction.tex
Execution control frameworks, such as SGX-Step~\cite{DBLP:conf/sosp/BulckPS17} and SEV-Step~\cite{sev-step} allow attackers to interrupt Trusted Execution Environments (TEEs) deterministically after every single instruction, a technique known as Deterministic Single-Stepping (DSS). DSS has given rise to several attack classes. In addition to amplifying microarchitectural leakage attacks, DSS has been an essential component in interrupt counting attacks~\cite{DBLP:conf/uss/MoghimiBHPS20, DBLP:journals/tches/AldayaB20, DBLP:conf/ccs/BulckOMAGP19, DBLP:journals/compsec/KimJPJKCK19}, a term coined by Constable et al.~\cite {aex-notify} to describe attacks that count interrupts to leak secrets protected by enclaves. Under DSS, interrupt counting becomes functionally equivalent to counting instructions. DSS is primarily enabled by the attacker priming microarchitectural behaviour, extending the execution time of instructions to the point where even inherently variable interrupt arrival times reliably fall within each instruction's execution window.

In response to interrupt-based stepping attacks, Intel has deployed AEX-Notify~\cite{aex-notify}, specifically designed to counter deterministic single-stepping on Intel SGX. AEX-Notify modifies enclave reentry, intending to prevent the attacker from extending instruction execution windows and from inferring interrupt landing points. To achieve this, AEX-Notify prefetches the working set of the next enclave instruction to ensure that data and code are correctly included in caches and TLBs, therefore reducing the instruction execution window. It further implements obfuscated forward progress, reducing an attacker's ability to distinguish between interrupts that occur during the prefetching mitigation and those that occur during the execution of instructions after the mitigation. By combining these techniques, AEX-Notify aims to prevent deterministic single-stepping and related attacks. 
Intel further implemented a fundamentally different mitigation against DSS in Intel TDX. However, this solution was shown to be insufficient in mitigating deterministic single steps~\cite{tdx-down}, with the authors noting that replicating the AEX-Notify design within Intel TDX could be a more viable defense strategy. AEX-Notify can therefore be seen as state-of-the-art mitigation against instruction stepping and interrupt counting attacks, and thus its design is potentially relevant beyond Intel SGX.

We argue that, in the context of interrupt counting attacks against SGX, focusing solely on preventing deterministic single-stepping may be misguided. We introduce AEX-NStep, the first practical interrupt counting attacks against Intel SGX since the deployment of AEX-Notify. Our attacks do not depend on deterministic single-stepping; instead, they are based on probabilistic observations. While the possibility of probabilistic attacks was discussed in~\cite{aex-notify}, these attacks were considered impractical, as they would require breaking AEX-Notify's obfuscated forward progress guarantee.

The key contribution of AEX-NStep is in introducing novel attack techniques and in showing that probabilistic interrupt counting attacks are practical. To achieve these attacks, we developed several new techniques: We concurrently run an attacker thread on the enclave core using SMT, creating contention between the two threads. This allows us to fingerprint the AEX-Notify mitigation by sampling performance counters, which measure the contention experienced by the attacker thread. This enables us to indirectly fingerprint the progress of the AEX-Notify mitigation, breaking AEX-Notify's obfuscated forward progress guarantee. We slow down the enclave application resumption point (EARP), i.e, the first instruction executed after the AEX-Notify mitigation, by disabling caches on the physical enclave core, and use Inter-Processor Interrupts (IPIs), which are fired from the concurrent attacker thread after enclave re-entry. This contrasts with the conventional APIC timer interrupts programmed before enclave re-entry, used in SGX-Step~\cite{DBLP:conf/sosp/BulckPS17}. Our approach ensures that fewer, potentially high-variance, instructions are executed between the point where the interrupt is programmed and when it is triggered, reducing the variability of interrupt arrival times and further contributing to interrupts landing within a close window to the EARP.

These techniques allow us to implement two novel and practical interrupt counting attacks: Probabilistic Single-Stepping (PSS) and Lower-Bounded Multi-Stepping (LBMS), neither relying on deterministic single-stepping.
In PSS, we attempt to approximate single-stepping. Although we cannot guarantee the execution of exactly one instruction between interrupts, we ensure that the number of instructions between subsequent interrupts is minimized, typically within a couple of instructions. This allows us to detect subtle instruction differences in conditional branches through repeated execution and subsequent averaging.
In LBMS, we lower-bound interrupt arrival times, such that conditional branches of different lengths result in deterministically distinct interrupt counts within a single trace. In both attacks, AEX-Notify is deployed, which prevents deterministic single-stepping. 

\taggedpara{Implications} Our work highlights the limitations of AEX-Notify. We show that, contrary to the AEX-Notify adversary model~\cite{aex-notify}, AEX-Notify is not suitable for SMT-enabled environments. We use LBMS to leak cryptographic keys from AEX-Notify-enabled SGX enclaves by exploiting nonce truncations during the ECDSA signature generation process, an attack previously shown in~\cite{tdx-down}. We leverage PSS to successfully execute the \texttt{memcmp()} attack on AEX-Notify-enabled enclaves; this attack was originally introduced in SGX-Step~\cite{vanbulck_sgxstep_memcmp}. Evaluating our PSS attack on a benchmark similar to AEX-Notify's, we show that attackers can single-step single, five-cycle memory-dependent instructions with a probability of 45\%.
Our results therefore show that, despite seemingly protecting against deterministic single stepping, AEX-Notify does not provide sufficient protection against interrupt counting attacks. These attacks will persist as long as attackers can exploit microarchitectural leakages to infer enclave execution progress and control interrupt timing during sensitive computations. Future defenses will need to directly address this capability to mitigate such attacks.

\taggedpara{Contributions}
\begin{itemize}
    \item We introduce AEX-NStep, the first practical interrupt counting attacks on Intel SGX since the deployment of AEX-Notify.
    \item We clarify the limits of AEX-Notify and demonstrate that, contrary to its adversary model, this mitigation does not protect against interrupt counting attacks in SMT-enabled environments. We notably show that the obfuscated forward progress guarantee does not hold.
    \item We show that deterministic single-stepping is not strictly required for interrupt counting attacks to be practical when considering AEX-Notify-enabled SGX enclaves. 
    \item Using our probabilistic interrupt counting attacks, we break \texttt{memcmp} ~\cite{vanbulck_sgxstep_memcmp} and leak an ECDSA private key on an AEX-Notify enabled SGX enclave. Prior to this work, interrupt counting attacks on these targets would have been considered to be prevented on Intel SGX by the deployment of AEX-Notify.   
\end{itemize}

\taggedpara{Responsible Disclosure} In line with the responsible disclosure guidelines in the CFP~\cite{IEEE_SP2026_EthicalDisclosure}, we notified Intel of our findings at the time of submission.

\taggedpara{Open Science} We publish all related code artifacts on Zenodo~\cite{dutly2025aexnstep} and Github~\cite{dutly2025github}. We hope that these artifacts will enable future research into AEX-Notify.

%% file: content/02-background.tex
\subsection{SGX-Step}

\input{content/background/background-sgx}

\input{content/background/background-sgx-step}

\subsection{AEX-Notify}
\input{content/background/background-aex-notify}

%% file: content/background/background-sgx.tex
Intel SGX enables the execution of processes that cannot be introspected by the operating system, which is considered untrusted under the SGX threat model. However, SGX relies on the operating system to manage memory. Consequently, it has full access to enclave page table entries (PTEs), including the unconstrained ability of setting PTE attributes and permissions, such as marking pages as non-executable (NX). Furthermore, the operating system is responsible for handling all asynchronous events that occur during enclave runtime, which can, for example, be triggered by accessing a protected page, executing an illegal instruction, or the arrival of an interrupt. In such cases, an Asynchronous Enclave Exit (AEX) occurs: The current CPU state is stored, and control is transferred to the operating system to handle the exception. Afterward, the \texttt{ERESUME} instruction is invoked to resume the enclave, which restores the CPU context. If enclave cooperation is needed, the OS can jump to a predefined enclave handler using \texttt{EENTER}. After handling the exception, the enclave exits via \texttt{EEXIT} and the OS resumes execution at the EARP using \texttt{ERESUME}.

\input{content/background/aex-notify-tikz}

%% file: content/background/aex-notify-tikz.tex
\begin{figure*}[ht]  %
\centering
\definecolor{pastelred}{RGB}{251, 206, 204}%
\definecolor{darkred}{RGB}{184, 84, 80}%
\usetikzlibrary{fit}
\resizebox{\textwidth}{!}{ %
\begin{tikzpicture}[node distance=0.8, auto]

    \node (sp) [input] {$s_p$};
    \node (dp) [input, right=of sp] {$d_p$};
    \node (cp) [input, right=of dp] {$c_p$};
    
    \node (start) [startstop, below=1.5cm of dp] {mitigation start};
    \node (decision) [decision, right=of start, fill=white] {Interrupted?};
    \node (process2) [process, above right=of decision, fill=white] {Restore $s_p,d_p,c_p$};
    \node (process3) [process, right=of decision, fill=white] {Save $s_p,d_p,c_p$};
    \node (stop) [process, right=of process2, draw= darkred, fill=pastelred] {\texttt{AEXNOTIFY=1}};
    \node (permissions) [process, right=of stop, fill=white] {\shortstack{Check PTE attributes \\ $s_p,d_p,c_p$ }};
    \node (warmup) [process, right=of permissions, fill=white] {\shortstack{Cache/TLB warmup\\$s_p,d_p,c_p$}};
    \node (random) [decision, right=of warmup, fill=white] {$r=1$?};
    \node (delay) [process, below=of random, fill=white] {20 NOP delay};
    \node (end) [startstop, right=of random] {jmp EARP};
\node[draw, dashed, darkred, rounded corners, 
      inner sep=0.3cm, label={[anchor=north west]north west:Atomic (\texttt{AEX\_NOTIFY=1})}] 
      (groupbox) [fit=(permissions) (warmup) (random) (delay) (end)] {};

    \draw [arrow] (sp) -- (start);
    \draw [arrow] (dp) -- (start);
    \draw [arrow] (cp) -- (start);
    \draw [arrow] (start) -- (decision);
    \draw [arrow] (decision) -- node[near start, above] {Yes} (process2);
    \draw [arrow] (decision) -- node[near start, below] {No} (process3);
    \draw [arrow] (process2) -- (stop);
    \draw [arrow] (process3) -- (stop);
    \draw [arrow] (stop) -- (permissions);
    \draw [arrow] (permissions) -- (warmup);
    \draw [arrow] (random) -- node[left] {Yes}(delay);
    \draw [arrow] (warmup) -- (random);
    \draw [arrow] (random) --  node[above] {No}(end);
    \draw [arrow] (delay) --(end);

\end{tikzpicture}
} %
\caption{Simplified control flow of the AEX-Notify mitigation stage II, which ensures that PTE attributes will not cause any faults (hence preventing zero-stepping) and that EARP will execute as fast as possible by priming its code ($c_p$), data ($d_p$), and stack ($s_p$) pages. If a random bit $r$ (passed implicitly as part of $c_p$ )is set to 1, additional random cycles are executed. Any interrupts during the atomic part will cause a roll-back of the previous $s_p$, $d_p$, and $s_p$ parameters, effectively restarting the mitigation.}
\label{fig:aex-flow}
\end{figure*}

%% file: content/background/background-sgx-step.tex
SGX-Step~\cite{DBLP:conf/sosp/BulckPS17} is an offensive research framework that provides fine-grained execution control over SGX enclaves:  While previous attacks were limited to tracking coarse enclave progress between page boundaries~\cite{xu2015controlled, DBLP:conf/uss/BulckWKPS17}, interrupting every enclave instruction (single-stepping) allows attackers to count intra-page enclave progress on an instruction-level granularity. Single-stepping and zero-stepping, which repeatedly faults the same instruction before retirement, have also been used to amplify microarchitectural leakages~\cite{DBLP:conf/uss/BorrelloKSLG022, DBLP:conf/uss/Moghimi23, DBLP:conf/ccs/0001LMBS0G19}.

Single-stepping is enabled by extending the execution window of each instruction to the point where it can be interrupted by carefully timed, but coarse-grained interrupts. By clearing the accessed bit of enclave page table entries before enclave re-entry, SGX-Step forces the next enclave instruction to trigger a CPU microcode-assisted path that sets the accessed bit to one, considerably lengthening its execution window~\cite{aex-notify}. 
If an interrupt occurs too early, the accessed bit will not be set, allowing the attacker to identify whether progress was made deterministically. By filtering zero-steps, SGX-Step achieves reliable single-stepping.

%% file: content/background/background-aex-notify.tex
AEX-Notify is an opt-in hardware-software co-design mitigation that makes SGX enclaves interrupt-aware and aims to prevent deterministic single-stepping (DSS), and zero-stepping (repeatedly faulting the same instruction). DSS attacks are prevented by priming the CPU and microarchitectural state so that the next enclave instruction after resuming from an asynchronous exit (the Enclave Application Resumption Point, or EARP) executes quickly, making it harder to interrupt. Zero-Stepping is prevented by verifying the EARP code and data PTE attributes. AEX-Notify can be activated as an opt-in mitigation when creating SGX enclaves. Enclaves can then enable or disable AEX-Notify at runtime by modifying the current thread control structure (TCS).

\taggedpara{First Mitigation Stage} The AEX-Notify mitigation is divided into two stages. The first stage decompiles the EARP to identify any data operands that may need pre-fetching.  AEX-Notify is not enabled in the TCS of the first stage, which can, as such, be single-stepped. Since single-stepping amplifies any leakages, special care is taken to ensure that this decompilation is constant-time and does not leak any information about the EARP. The first stage invokes the second mitigation stage with pointers to data, code, and stack addresses used by the EARP.

\taggedpara{Second Mitigation Stage} The second mitigation stage, visualized in Fig.~\ref{fig:aex-flow}, ensures that the working set of the EARP is primed. It is implemented in a carefully aligned assembly stub to prevent any observable cross-page accesses from being used to infer the progress of the mitigation.  After it verifies that the last AEX did not occur during the second mitigation stage, it enables the AEX-Notify bit in the current TCS, and it verifies that the EARP code and data pages have valid permissions. This ensures that the EARP will not directly fault and cause a zero-step. Next, it repeatedly accesses the code, data, and stack pages in a tight loop to ensure appropriate entries are available in caches and TLBs. This ensures fast EARP execution and is designed to mitigate single-stepping. Lastly, in order to make the execution time of the mitigation unpredictable, a fixed delay is executed if a random bit (an implicit parameter in the code pointer) is set to 1. This delay is implemented as a series of \texttt{lea rsp,[rsp]} instructions, henceforth referred to as the NOP delay. 

\taggedpara{Changes to \texttt{ERESUME}} If AEX-Notify is enabled, the \texttt{ERESUME} instruction behaves akin to \texttt{EENTER}. It causes the control flow to jump to the AEX-Notify mitigation stub instead of directly resuming the execution of the EARP. A new instruction (\texttt{EDECCSSA}) enables continuous control flow between the mitigation stage and the EARP, eliminating the need to exit the enclave and reenter it via \texttt{ERESUME}.

%% file: content/03-attack-overview.tex
\section{Implementing PSS and LBMS on Hardened AEX-Notify Enclaves} 
\label{sec:details}

\begin{figure*}
    \centering
    \includegraphics[width=\linewidth]{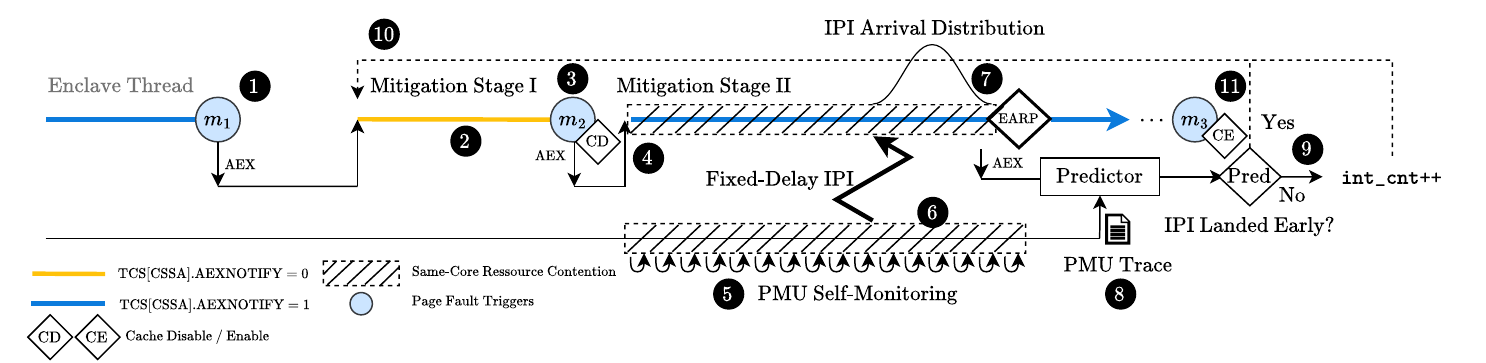}
    \caption{Simplified probabilistic single-stepping (PSS) attack flow: The attacker's goal is to approximate the behavior of deterministic single-stepping on a section of enclave code, delimited by page fault markers $m_1$ and $m_3$ (\circlednum{1}, \circlednum{11}). Since the first AEX-Notify mitigation stage (\circlednum{2}) runs without setting the AEX-Notify bit to one in the current TCS, an attacker can reliably single-step this portion of code, advancing the enclave until $m_2$, the last instruction before AEX-Notify is enabled (\circlednum{3}). This is consistent with AEX-Notify's threat model~\cite{aex-notify}. Before resuming the enclave's execution (\circlednum{4}), the attacker starts a concurrent thread running on the same physical core as the enclave. The enclave's mitigation stage II code shares microarchitectural resources with the concurrent attack thread, which can monitor this contention by sampling \texttt{UOPS\_EXECUTED.CORE\_CYCLES\_NONE} counters (\circlednum{5}). After a fixed delay, the attacker thread fires an IPI (\circlednum{6}), which interrupts the enclave within a given arrival distribution (\circlednum{7}). The interrupt will trigger an AEX, at which point the attacker will be able to classify the performance counter trace to determine whether the interrupt landed during the mitigation (\circlednum{8}, \circlednum{9}). By repeating this process (\circlednum{10}), the attacker will eventually land interrupts within a couple of instructions of the EARP, achieving probabilistic single-stepping.}
    \label{fig:attack-flow}
\end{figure*}

\subsection{Probabilistic Single-Stepping (PSS)}
\label{sec:details-pss}
PSS aims to maximize the likelihood of single-stepping. Unfortunately, interrupt arrival times have a significant variance of several hundred cycles~\cite{aex-notify}.  A naive approach to probabilistic single-stepping would be to align the mean of the interrupt arrival distribution with the EARP execution window. However, this would result in half of the interrupts landing during the AEX-Notify Stage II mitigation, with the other half potentially overshooting the EARP by a large margin. 

\taggedpara{Reducing over-shooting} To reduce the number of over-shot instructions, an attacker can shift the interrupt arrival distribution to line up its right tail with the EARP execution window by firing interrupts earlier. Now, most interrupts land in the stage II mitigation, and a few interrupts land during (or shortly after) the execution of the EARP. Extending this execution time further increases the likelihood that these interrupts will land within the first few instructions of the enclave. We achieve this by disabling caching in the enclave core between page faults, rendering the cache warm-up loop of the stage II mitigation ineffective\footnote{In our practical benchmarks and attacks (Sections 5 and 6), disabling caches incurs a 275-500x performance penalty on the SGX enclave, depending on the enclave code.}.

However, disabling caches increases the variability of APIC timer interrupts programmed before enclave reentry, due to 
increased variance caused by DRAM operations between arming the timer and the interrupt landing, as noted in the security analysis of AEX-Notify~\cite{aex-notify}. To counteract this, we rely on inter-processor interrupts (IPIs), which are fired after enclave reentry from a concurrent attacker thread. This minimizes the impact of high-variability memory operations performed at the start of the mitigation.

\taggedpara{Breaking Obfuscated Forward Progress} \label{breaking-obf} For successful interrupt counting attacks, interrupts landing during the mitigation should not add to the interrupt count. However, AEX-Notify's obfuscated forward progress guarantee prevents attackers from determining whether an interrupt landed during or after the mitigation using conventional methods.

Instead, we use the co-located attacker thread to fingerprint the progress of the mitigation. As it is running on the same core as the enclave, the stage II mitigation code shares a CPU cycle budget with the concurrent attacker thread. This generates contention between the two threads.
The attacker can measure this contention by sampling the core-specific \texttt{UOPS\_EXECUTED.CORE\_CYCLES\_NONE} performance counter that is, in contrast to thread-specific counters, not affected by SGX mitigations concerning performance counters (ASCI~\cite{intel_sgx_prog_ref}). This counter reports the number of cycles for which no $\mu$-ops were executed. The enclave and the attacker thread both execute memory-dependent operations in tight succession: The enclave reads stack, code, and potentially data pages used by the EARP in a tight loop, while the attacker concurrently samples performance counters and writes these to an in-memory buffer. Since caching is disabled, concurrent memory pressure results in  CPU cycles for which no $\mu$-ops are executed. While the attacker sampling loop always performs the same sequence of instructions, the enclave progresses through the stage II mitigation. Therefore, the performance counters collected by the attacker thread fingerprint the stage II mitigation.

We use a high-frequency performance counter sampling loop by directly sampling counters via the \texttt{rdpmc} instruction, which can be executed at any privilege level by setting the PCE flag~\cite{RDPMC}. The sampling loop gathers counters from when the enclave resumes stage II mitigation until shortly after firing the IPI, resulting in 120 sampled values per interrupt. Collected traces exhibit recognizable patterns which can be mapped to events occurring during the mitigation (Fig.~\ref{fig:pmctrace}). While the number of sampled performance counters remains fixed, the mitigation's progress will vary depending on how physical core cycles are shared between the enclave thread and the attacker sampling thread, as well as when the IPI is triggered. 
This causes a slight shift in trace features, which we classify using a random forest model. To train the model, we run a debug enclave and generate labeled traces for each interrupt, according to whether the interrupt occurred during the mitigation ($n<0$), zero-stepped the EARP ($n=0$), or single- or multi-stepped the EARP ($n\geq 1$).

\taggedpara{Dealing with randomized NOPs} As explained in Fig.~\ref{fig:aex-flow}, AEX-Notify executes a fixed amount of NOP instructions at the end of the mitigation with probability $p=0.5$. However, this random bit is sampled before the critical AEX-Notify assembly stub and passed as an implicit parameter as part of the code page tickle parameter. Given that this parameter is backed up and restored if an interrupt lands during the mitigation, the bit controlling the execution of the random NOP instructions will also be restored. As such, the behavior of the NOP slide is identical between AEXs caused by interrupting the mitigation. This allows attackers to choose a conservative delay before firing the IPI, calibrating them to hit the EARP when no random NOPs are executed. If the trace classifier detects many interrupts that fail to clear the mitigation, the attacker infers that these are repeatedly landing in the NOP slide. The attacker can counter this by slightly delaying the IPI to clear the slide. When classifying traces, interrupts that land during mitigation and interrupts zero-stepping the EARP must be handled differently, as a zero-step will cause a new random bit $r$ to be sampled, potentially altering the behavior of the NOP slide.

An attacker's ability to determine whether interrupts landed early completely negates any benefit of adding a randomized delay before executing the EARP. Even if the random bit controlling the randomized delay is re-sampled every time the mitigation runs, an attacker could choose conservative timings for the IPIs (assuming that no delay is executed). On average, this would result in half of the interrupts being fired in no-delay cases. The other half would land in the mitigation and be discarded by the attacker after being classified as early interrupts.

Fig.~\ref{fig:attack-flow} depicts the PSS attack flow, including the monitoring phase and firing the IPI from the concurrent attacker thread.

\subsection{Lower-Bounded-Multi-Stepping (LBMS)}
\label{sec:details-lbms}
The goal of LBMS is to fire an interrupt after at least $n$ enclave instructions have executed. In contrast to PSS, implementing LBMS poses fewer technical challenges. First, LBMS attacks do not require breaking obfuscated forward progress: As IPI delays are calibrated such that the IPI triggers earliest after $n$ enclave instructions, interrupts never land during the mitigation. The IPI lower bound can be calibrated under the assumption that the mitigation's random delay never executes. Hence, the random NOP delay does not affect attackers in the LBMS use case. The attacker still relies on disabling caches during the critical code section to reduce the execution window of enclave instructions and utilizes an IPI fired from a concurrent thread to minimize the variance of interrupt arrivals.

%% file: content/04-attack-analysis.tex
We quantify the behavior of our probabilistic single-stepping (PSS) and lower-bounded multi-stepping (LBMS) primitives. For PSS, particular focus is placed on evaluating our ability to classify interrupts depending on whether they landed in the mitigation, zero-stepped, or $n\geq1$-stepped the EARP. We then quantify the probability of observing single steps given our PSS primitive and evaluate the average attacker success rate of our PSS primitive in distinguishing subtle control flow differences. For LBMS, we assess the impact of branch deltas on the detection rate. 

\subsection{Experimental Setup}

We run all experiments on an Ubuntu 22.04.1 machine with a 6.8.0-38-generic kernel, powered by an Intel i9-9900K CPU (microcode \texttt{0xf8}) running at a fixed frequency of 3.60 GHz with disabled hardware power states. We use the SGX driver provided by the 6.8.0 Linux kernel and the Linux SGX SDK Gold Release 2.23. We run all of our experiments with AEX-Notify enabled. Enclaves are spawned in debug or production mode depending on the experiment. As was done for the security analysis of AEX-Notify~\cite{aex-notify}, we modify the SDK to induce a page fault on the last instruction before AEX-Notify is enabled in the stage II mitigation. This simulates an attacker that can single-step non-AEX-Notify protected enclave threads, following the attacker capabilities demonstrated by prior research under the SGX threat model. We assume that the attacker has access to conventional 4KiB controlled side channels~\cite{DBLP:conf/uss/BulckWKPS17, xu2015controlled} to isolate points of interest in an enclave before running PSS/LMBS attacks. We do, in particular, not consider any mitigations that would limit this spatial resolution~\cite{DBLP:conf/ccs/ShindeCNS16, vanoverloop2025tlblur}.

\taggedpara{Preventing superfluous interrupts} We ensure that the target enclave is only interrupted by attacker-originated IPIs by isolating the enclave core and offloading kernel housekeeping threads and default \texttt{irq} affinities to other cores. We use a FIFO real-time scheduling policy with unbounded runtime restrictions for both the enclave and attacker thread.

\subsection{Filtering Early-Interrupts}
\label{modeleval}

We provide a quantitative evaluation of our ability to successfully distinguish between interrupts that single- or multi-step the EARP ($n\geq 1$ interrupts), those that zero-step ($n=0$), and those landing during the mitigation ($n<0$).
\begin{figure}
    \centering
    \includegraphics[width=\linewidth]{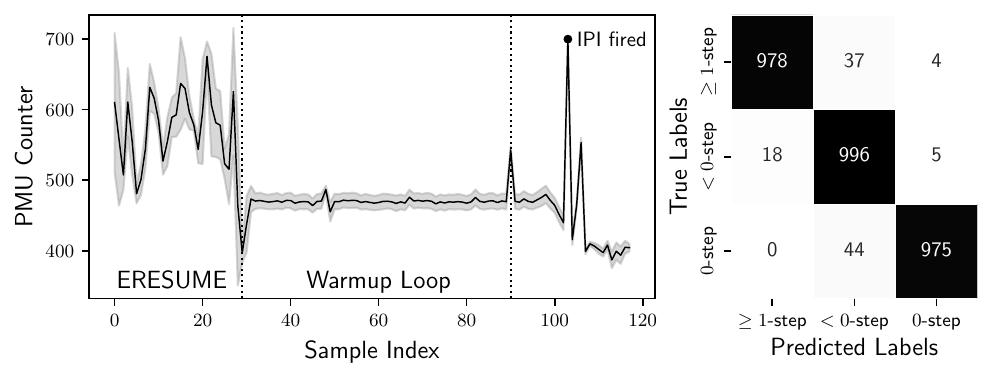}
    \caption{Left: Mean \texttt{UOPS\_EXECUTED.CORE\_CYCLES\_NONE} rate of change ($N=684$), sampled on the attacker thread during enclave re-entry. The trace fingerprints the AEX-Notify stage II mitigation, revealing forward progress leakage: While the number of sampled counters is always fixed, the progress made by the mitigation varies depending on hyper-threading cycle allocations between the enclave and the attacker thread. This causes shifts in trace features, which can be classified to distinguish interrupts based on their arrival locations.}
    \label{fig:pmctrace}
\end{figure}

\taggedpara{Offline Phase} We gather $4,099$ labeled performance counter traces (cf. Section~\ref{breaking-obf}) for each of the three classes by running the target enclave in debug mode and inspecting the enclave return instruction pointer after each interrupt. We gather traces for an enclave executing a uniform sequence of memory-dependent \texttt{addl} instructions and then evaluate the transferability of our trained model by classifying interrupts on an enclave executing a uniform sequence of \texttt{nop} instructions. These instruction choices are selected to evaluate our resulting stepping rate, and will be further elaborated upon in Section \ref{quantifying-pss}. By choosing \texttt{nop} instructions, we provide a lower bound on the resolution of our attack as they only require $0.25$ cycles to execute. We use five-cycle memory-dependent \texttt{addl} instructions to upper-bound the resolution of our attack: Five cycles exceed the best-case cycle requirements for most common instructions and will, in practice, require many more cycles to execute, given the reliance on a memory operand~\cite{fog_instruction_tables}.

\taggedpara{Model Training} We use 20\% of our training set as a test set ($\sim 819$ samples per class), allowing the training of a random forest classifier on the remaining traces. We achieve an F1 score (harmonic mean between precision and recall) of 0.95 for interrupts landing during the mitigation ($n < 0$), an F1 score of $0.99$ for zero-stepping interrupts ($n=0$), and an F1 score of $0.96$ for interrupts resulting in $\geq1$-steps. A confusion matrix for test set predictions is provided in Fig.~\ref{fig:pmctrace} (right).

\taggedpara{Practical Model Evaluation} During PSS attacks, most interrupts will land during the mitigation. This imbalance increases the probability that early interrupts will be misclassified into other classes. To consider this and evaluate our classifier online, we predict interrupt classes when performing a PSS attack on a uniform sequence of 500 \texttt{addl} instructions. For $n \geq 1$ interrupts, we report precision and recall values of $0.94$ and $0.89$, respectively.

\taggedpara{Model Transferability} We test the transferability of our approach by training on traces collected from a sequence of uniform \texttt{addl} instructions and applying it to predict interrupt locations when benchmarking on a sequence of 500 uniform \texttt{nop} instructions. We report a $\geq 1$-step prediction precision of $0.56$ and a corresponding recall of $0.82$. We attribute the degradation in predictive model performance to differing cycle requirements between instruction types. Retraining on the new target yields a precision of $0.86$ and a recall of $0.95$ when classifying $n \geq 1$ interrupts. We therefore recommend collecting and training on target-specific traces to improve prediction accuracy, which is permissible under the SGX threat model.

\taggedpara{Effects of Mitigation NOPs} Given the non-deterministic behavior of our classifier, we explored whether the random execution of the mitigation NOP delay contributes to this variability. To test this, we analyzed predictions for interrupts occurring during the NOP slide and observed a precision of 1.0 and a recall of 0.98 for the [addl] benchmark ($N = 1,165$). These results suggest that the probabilistic behavior of our attack is unlikely to be attributed to the execution of the NOP delay.

\taggedpara{Comparison with AEX-Notify} Our evaluation shows that, in contrast to the assumptions made by the AEX-Notify paper, it is possible to differentiate interrupts according to their arrival locations. However, we note that while our method achieves good predictive performance, it relies on an imperfect classifier, which poses a further hurdle should attackers attempt to convert PSS into a deterministic single-stepping primitive.

\subsection{PSS Stepping Rate}
\label{quantifying-pss}
We quantify our probabilistic stepping rate, i.e., the probability of $n$-stepping for different values of $n$, by running an enclave in debug mode and repeatedly interrupting a sequence of 500 uniform \texttt{addl} or \texttt{nop} instructions. Running the enclave in debug mode enables us to measure the progress (if any) an enclave has made between interrupts. While this will allow us to characterize the behavior of our PSS primitive and draw comparisons with AEX-Notify, we note that attackers do not need to know or estimate the probabilistic stepping distributions when performing attacks on production enclaves. 

For each interrupt classified by our model as a $ \geq 1$ interrupt, we examine the enclave return instruction pointer (eRIP) to determine how many instructions were executed since the last $ \geq 1$ interrupt. We separately report false positives, i.e., early (mitigation) interrupts or zero-steps misclassified as $ \geq 1$ interrupts.

\begin{figure}
    \centering
    \includegraphics[width=\linewidth]{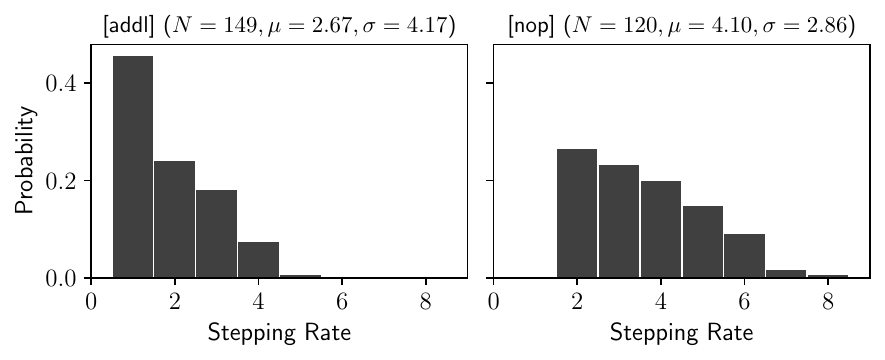}
    \caption{Probabilistic stepping rates for sequences of 500 \texttt{addl} (left) and \texttt{nop} (right) instructions, measured on a debug enclave by observing the enclave's instruction pointer after each post-mitigation interrupt. For memory-dependent \texttt{addl} instructions, around 45\% of all correctly classified post-mitigation interrupts result in single-stepping. We observe no single-stepping for \texttt{nop} instructions, and the peak of the stepping distribution is shifted to two-stepping (25\%).}
    \label{fig:steping-rate}
\end{figure}
We achieve a single-stepping probability of $45\%$ when tracing \texttt{addl} instructions. We do not observe any single steps for \texttt{nop} instructions, although the two-stepping probability rises to around 25\%. The exact probability distributions are provided in Fig.~\ref{fig:steping-rate}. We observed a total of 28 false positives for \texttt{addl} instructions and \texttt{6} false positives for \texttt{nop} instructions. The cases mostly result from interrupts landing in the mitigation and being misclassified as $n\geq 1$ interrupts.  
Given the high number of interrupts landing in the mitigation, the number of false positives is relatively small. For the \texttt{nop} benchmark, $1741$ interrupts landed in the mitigation, representing around $90\%$ of all interrupts fired during the PSS attack.

\taggedpara{Comparison with AEX-Notify} Our results highlight a non-negligible single-stepping rate for \texttt{addl} instructions. Our stepping rates should not be possible under the guarantees given by AEX-Notify, most notably because the AEX-Notify paper assumes that attackers cannot distinguish where interrupts land and whether they resulted in a $n\geq 1$-step.

\subsection{PSS: Evaluating Attacker Success Rates}

We now evaluate our PSS primitive on a synthetic benchmark to characterize an attacker's success rates in detecting subtle control flow imbalances for different sample counts per branch. We construct a benchmark consisting of a $\Delta$-unbalanced branch which depends on a secret bit $s$, where the short branch is taken if the secret matches an attacker's guess $g$, and the longer branch is taken if the attacker's guess is incorrect (Listing~\ref{list:benchmark}). The attacker's goal is to leak the secret bit $s$. To make the prediction, the attacker samples $k$ traces for each guess $g\in\{0,1\}$ using the PSS primitive, collecting several observed interrupts for each trace. Each fired interrupt will land in the mitigation, zero-step the EARP, or result in a $n \geq 1$- step. This process begins after the first page fault marker and ends at the second page fault marker. The enclave only makes progress if interrupts result in a $n \geq 1$-step. By only counting $n  \geq 1$ interrupts and filtering all other interrupts (Section~\ref{modeleval}), an attacker can then compute the mean $n \geq 1$ interrupt count for sets of $k$ traces collected per guess and predict $s=g$ where $g$ corresponds to the guess with the smaller mean number of interrupts among $k$ traces. We assume that any other direct or indirect observations cannot differentiate between correct and incorrect guesses. In practice, an attacker may need to execute such an attack for every bit of a cryptographic key, only being able to verify correctness after attempting to leak every bit of the key. We report the average attack success rate over 50 trials for various sample values $k$ and different branch imbalances $\Delta$. All experiments in this section are executed on a production SGX-Enclave with AEX-Notify enabled.

\begin{figure}[tbp]
\renewcommand{\figurename}{Listing}
    \centering
\begin{lstlisting}[language=C]
mwrite(&start_tracing); // PF trigger, NX
if (s == attacker_guess){
    __asm__ __volatile__ (
        " .rept " STR(10) "\n"
        " nop\n"
        " .endr\n"
    );
}else{
    __asm__ __volatile__ (
        " .rept " STR(10) + DELTA "\n"
        " nop\n"
        " .endr\n"
    );
}
mwrite(&stop_tracing); // PF trigger, NX
\end{lstlisting}
\caption{{\captionsize{Our PSS benchmark measures attacker success rate for leaking a secret bit $s$ based on various branch imbalances and varying numbers of PSS interrupt traces collected for each guess. The benchmark consists of a $\Delta-$unbalanced branch between two page fault markers, used to start and stop collecting traces. Under the probabilistic single-stepping model, the average number of interrupts observed for correct attacker guesses will be smaller than the average interrupt count for incorrect guesses.}}}
\label{list:benchmark}
\end{figure}

\begin{figure}
\centering
\includegraphics[width=\linewidth]{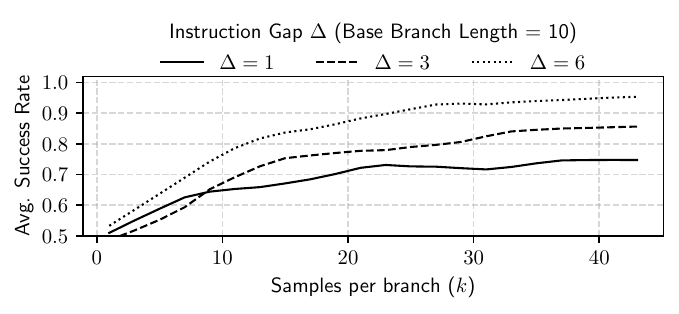}
\caption{PSS average attacker success on our synthetic benchmark (Listing~\ref{list:benchmark}). Measured over 50 trials when discriminating secret-dependent $\Delta$-unbalanced branches for different sample counts per branch. Values are smoothed to account for residual fluctuations caused by small sample and trial counts. The trend indicates that higher delta values result in smaller sample sizes required to achieve a target average success rate.}
\label{fig:encl_run}
\end{figure}

\taggedpara{Results} Our results are provided in Fig.~\ref{fig:encl_run}. Generally, higher sample counts per guess result in higher success rates. Lower instruction deltas between unbalanced branches decrease an attacker's success rate, considering a fixed sample size, compared to higher instruction deltas. After gathering $40$ samples per guess, the average attacker's success rate in correctly deducing the secret $s$ lies at around $70\%$ for a one-instruction difference between branches, at $85\%$ for a three-instruction difference, and $95\%$ for a six-instruction difference. Given that our unbalanced branches mainly consist of \texttt{nop} instructions, these results are lower bounds for the resolution achievable with probabilistic single-stepping.

\subsection{LBMS: Impact of Instruction Differences}
\label{charac-lbms}
During LBMS, the attacker lowers bounds the IPI arrival distribution by the number of instructions in the shorter branch. Only traces during which the longer branch executes can contain an interrupt. However, not every trace of the longer branch leads to an interrupt: If the instruction delta between both branches is small, interrupts may fail to trigger before observing the boundary page AEX (Fig.~\ref{fig:lbms-biased}).

We reuse the code from Listing~\ref{list:benchmark} to measure the number of traces during which an interrupt occurred, over 1000 executions of the longer branch. Interrupts are calibrated such that the execution of the shorter branch does not lead to any observable interrupts. We run this benchmark for $\Delta \in \{2,4,6,8,16,32,64\}$ and average the number of observed interrupts for every delta over 10 runs.

\taggedpara{Results} Table~\ref{tab:interrupts} shows that the average number of traces with interrupts increases as delta grows. For ephemeral secrets, longer execution paths may result from rare events, like a biased nonce in ECDSA. LBMS is effective for code with either large instruction differences triggered by rare events (Section~\ref{section:tdx}) or small differences with moderate likelihood (Section~\ref{section:copycat}).

\begin{table}
\centering
\caption{LBMS traces with interrupt observed per 1000 traces taking the longer branch for varying $\Delta$ values.}
\begin{tabular}{c|cccccc}
$\Delta$ & 2 & 4 & 8 & 16 &32 &64
\\
\hline
Nr. of Traces & 63 & 155 & 431 & 483 &995 & 1000

\end{tabular}
\label{tab:interrupts}
\end{table}

%% file: content/05-exploitation.tex
We evaluate our PSS and LBMS attacks on practical targets. For PSS, we target one of the original \texttt{memcmp} proof-of-concepts shipped with SGX-Step, which leaks a secret string by exploiting a non-constant-time string comparison function~\cite{vanbulck_sgxstep_memcmp}.
For LBMS, we turn to targets that rely on ephemeral secrets. In particular, we target ECDSA signatures, where nonce biases can be exploited to recover the private signing key through lattice reduction attacks~\cite{DBLP:conf/fc/BreitnerH19}. We target two previously published vulnerabilities in the WolfSSL cryptographic library~\cite{DBLP:conf/uss/MoghimiBHPS20, tdx-down}.

\subsection{PSS: Replicating SGX-Step's \texttt{memcmp} Attack}
\label{replicating-memcmp}
SGX-Step's \texttt{memcmp} example is a non-constant time string comparison that checks attacker guesses against an enclave-protected secret. Based on a real-world SGX vulnerability~\cite{van2019tale}, the function exits early on length or character mismatches, as shown in Listing~\ref{list:memcmp}.

\taggedpara{Offline Phase} We train our interrupt classifier by repeatedly running the enclave in debug mode with a fixed, arbitrary secret. We gather a total of $90,455$ traces for each target class, namely interrupts landing during the mitigation, interrupts zero-stepping the EARP, and interrupts resulting in $ n \geq 1$-stepping. We train a random forest classifier on the collected traces, achieving a test set (20\% of all traces) F1 score of $0.91$ for $n\geq 1$ interrupts, $0.94$ for mitigation interrupts, and $0.96$ for zero-steps.

\taggedpara{Online Phase} We use our trained model to classify interrupts on a production enclave, collecting interrupt counts with PSS for various attacker guesses. To determine the secret's length, we input strings of lengths one to eight.  A matching length results in more instructions executing and a higher average interrupt count. We average over 250 traces per guess. Afterwards, we repeat the process to recover each character, testing 26 candidates per character (\texttt{A-Z}).

\taggedpara{Evaluation} The attack requires a total of $3,493,950$ interrupts to recover a six-character secret (\texttt{SECRET}), totaling eight hours of compute time. Using conventional deterministic single-stepping on a non-AEX-Notify SGX enclave, the same attack finishes in under a second and requires 73 interrupts. The difference in interrupt counts can be attributed to the high number of interrupts landing during the AEX-Notify mitigation and the averaging procedure required for each character. The authors of SGX-Step included an experiment~\cite{vanbulck_sgxstep_memcmp} to test whether simple timing side channels could also exploit the non-constant-time string comparison. Their measurements indicate that using simple TSC timing measurements yields no statistically significant timing difference when averaging over $100,000$ samples.

\begin{figure}
\renewcommand{\figurename}{Listing}
\begin{lstlisting}[language=C]
int my_memcmp(char *a, int a_len, char *b, int b_len){
    int i;
    if (a_len != b_len)
        return 0;
    for (i=0; i < a_len; i++){
        if (a[i] != b[i])
            return 0;
    }
    return 1;
}
\end{lstlisting}
    \caption{SGX-Step's non-constant time \texttt{memcmp} example~\cite{vanbulck_sgxstep_memcmp}, originally used to highlight the effectiveness of deterministic single stepping. The snippet exhibits early exit patterns for non-matching lengths and character comparisons, resulting in variable instruction counts depending on the input. The example is based on a real-world vulnerability in Intel SGX~\cite{van2019tale}.}
    \label{list:memcmp}

\end{figure}

\subsection{LBMS: Exploiting Nonce Truncations}
\label{section:tdx}
Nonces used during ECDSA signatures must conform to specific requirements; in particular, they must be smaller than the modulus $n$ of the chosen elliptic curve. Two methods can be used to achieve this: Rejection sampling and modular reduction-based truncation. As shown by~\cite{tdx-down}, the nonce truncation routine in wolfSSL 5.6.4 exhibits a non-constant time behavior for certain curves: For the curve \texttt{secp160r1}, the \texttt{do\_while} in the \texttt{\_sp\_div\_impl} routine will execute two iterations for particularly large nonces (15 most significant bits set to one) and at most one complete iteration otherwise (Listing~\ref{list:nonce_truncation}). By building a page-fault-based state machine, attackers can isolate the execution of this loop, resulting in a 52-instruction delta depending on whether the generated nonce exhibits the 15-bit bias or not~\cite{tdx-down}. While the generation of such large nonces is unlikely ($p=1.5 \cdot 10^{-5}$), the detection of 12 biased signatures is sufficient to leak the key. We assume that an attacker can isolate the execution of the target \texttt{do\_while} loop using a page fault state machine~\cite{tdx-down}, which we simulate by placing appropriate page fault markers before and after the outermost \texttt{for} loop, as shown in Listing~\ref{list:nonce_truncation}.

\begin{figure}
\renewcommand{\figurename}{Listing}
\begin{lstlisting}[language=C]
static int _sp_div_impl(sp_int* a, const sp_int* d, 
                        sp_int* r, sp_int* trial)
// start PF marker
for (i = a->used - 1; i >= d->used; i--) {
    do {
         for (j = d->used; j > 0; j--) {
             if (trial->dp[j] != a->dp[j + o]) {
                 break;
             }
         }
         if (trial->dp[j] > a->dp[j + o]) {
             t--;
         }
     }
     // 15-bit biases result in two do-while iterations
     while (trial->dp[j] > a->dp[j + o]);
    // (..)
}
// end PF marker
\end{lstlisting}
\caption{The \texttt{\_sp\_div\_impl} routine to truncate nonces for secp160r1 curves. As noted by~\cite{tdx-down}, the do-while loop is executed twice for a 15-bit biased nonce, resulting in a difference of about 50 instructions.}
\label{list:nonce_truncation}
\end{figure}

\taggedpara{LBMS Variations} LBMS is not limited to scenarios where attackers observe only zero or one interrupt. If the instruction gap between the two branches is large enough, attackers can calibrate interrupts so that any observed difference in interrupt count indicates which branch was executed, which facilitates the IPI delay calibration. For example, interrupts can be tuned so that the shorter branch produces at most one interrupt, while the longer branch produces at least two. We use this flexibility for the ECDSA nonce truncation attack.

\taggedpara{Offline Phase} To reduce the time required for calibrating IPIs, we modify the random number generation process of WolfSSL to ensure that every eighth signature is generated with a 15-bit biased nonce. We then continuously shift the point in time at which the IPI is fired until we observe a clear difference in the number of interrupts when signing a message with a biased nonce.  Given the larger instruction delta between the biased and non-biased case, we calibrate our delay such that at most one interrupt is observed for non-biased nonces and two interrupts for 15-bit biased nonces.

\taggedpara{Online Phase} We run the attack on a production enclave with AEX-Notify enabled, collecting signatures until we observe two interrupts for 16 of them. We did this using the stock random number generator in WolfSSL, reverting the probability of observing 15-bit biased nonces to $1.5 \cdot 10^{-5}$. Collecting 16 signatures with two observed interrupts required performing a total of $922,933$ signatures, which required five hours and 42 minutes of compute time. In contrast to~\cite{tdx-down}, we do not need an oracle that leaks the number of instructions executed between each interrupt.

\taggedpara{Lattice Reduction} As noted by~\cite{tdx-down}, a successful LLL-lattice reduction attack requires 12 biased signatures. We perform LLL reductions on random subsets of 12 from our 16 collected traces and recover the ECDSA private key after 2,726 reductions, completed in 2 minutes using parallelized reductions at a rate of 65 reductions per second.

\subsection{Limits of LBMS: Targeting Minimal Deltas}
\label{section:copycat}
We evaluate the limits of LBMS by targeting a victim with minimal secret-dependent branch deltas, specifically a WolfSSL vulnerability discovered by CopyCat~\cite{DBLP:conf/uss/MoghimiBHPS20}. ECDSA signatures involve computing the scalar multiplication $k \times G$ for a given nonce $k$ and generator $G$. In WolfSSL versions before 4.2.1, this computation utilizes a hardened double-and-add algorithm, designed to perform the same number of operations for each bit of $k$. However, CopyCat showed that the compiled binary still contained a variable number of instructions depending on whether the current bit was a leading zero bit (LZB). This difference leaks the bit length of the nonce $k$, enabling the extraction of the private signing key.
We were unable to reproduce the exact four instructions difference reported by CopyCat when we compiled the vulnerable WolfSSL version with \texttt{GCC 11.4.0}. We did, however, identify a two-instruction difference between \texttt{mp\_copy} function calls when dealing with leading zero bits (Listing~\ref{list:asmcopy}).
We target signatures with five leading zero bits, as they strike a good balance between occurrence probability ($p = 0.03$) and practicality for a subsequent lattice reduction attack, requiring only 34 biased signatures on secp160r1 curves when using a BKZ-based approach.

\noindent
\begin{figure}
\renewcommand{\figurename}{Listing}
\begin{minipage}[t]{0.48\columnwidth}
\begin{lstlisting}[basicstyle=\ttfamily\tiny]
call   167fe <mp_copy>
test   eax,eax
jne    48a9 <wc_ecc_mulmod_ex+0x758>
lea    r11,[r12+0x248]
mov    rax,QWORD PTR [rsp+0x68]
and    rax,rbx
mov    rdx,rax
mov    QWORD PTR [rsp+0x18],r13
lea    rax,[r13+0x248]
and    rax,QWORD PTR [rsp+0x30]
lea    r10,[rdx+rax*1]
mov    QWORD PTR [rsp+0x58],r11
mov    rsi,r11
mov    QWORD PTR [rsp+0x60],r10
mov    rdi,r10
call   167fe <mp_copy>
\end{lstlisting}
\end{minipage}
\hfill
\begin{minipage}[t]{0.48\columnwidth}
\begin{lstlisting}[basicstyle=\ttfamily\tiny]
call   167fe <mp_copy>
test   eax,eax
jne    48cc <wc_ecc_mulmod_ex+0x77b>
lea    r10,[r12+0x248]
mov    r14,QWORD PTR [rsp+0x68]
and    r14,rbx
mov    QWORD PTR [rsp+0x18],r13
lea    rax,[r13+0x248]
and    rax,QWORD PTR [rsp+0x38]
add    r14,rax
mov    QWORD PTR [rsp+0x60],r10
mov    rsi,r10
mov    rdi,r14
call   167fe <mp_copy>
\end{lstlisting}
\end{minipage}
\caption{Left: Instruction sequence between \texttt{mp\_copy} calls when processing a leading zero bit. Right: Instruction sequence between \texttt{mp\_copy} calls when processing bits starting from the first non-zero MSB. These sequences are part of the scalar multiplication routine used to compute $k\times G$. The 2-instruction difference leaks the bit length of the nonce $k$.}
\label{list:asmcopy}
\end{figure}

\taggedpara{Offline Phase} We design a page-fault-based state machine that isolates the relevant \texttt{mp\_copy} calls when the fifth most significant bit of the nonce is being processed. If the current nonce has at least five leading zero bits, 14 instructions execute between page fault boundaries at offsets starting with \texttt{0x47}. Otherwise, if a leading non-zero bit appears before the fifth most significant bit, 12 instructions execute between boundaries with offsets near \texttt{0x45}.
We run the target code in debug mode and trigger an IPI at each \texttt{mp\_copy} call pair, adjusting the IPI delay to only observe eRIP values with offsets starting with \texttt{0x47} (biased nonces). This calibration process revealed that most interrupts (75\%) result in eRIP values equal to the second \texttt{call mp\_copy} instruction, regardless of whether the current nonce is biased.

\taggedpara{Filtering \texttt{call} instructions} We use a timing side channel to deterministically filter cases where enclave return pointers correspond to the second \texttt{call} instruction. The AEX-Notify mitigation checks if the EARP is executable. By measuring the TSC delta before stage II mitigation and the page fault handler, we detect if the eRIP matches the \texttt{call} instruction. If so, an early page fault occurs during execution of the target page. If the interrupt happened before the call, no early fault occurs, and the cycle delta before the \texttt{mp\_copy} page fault is longer. eRIP values at \texttt{call} instructions yield a 5.7\% smaller TSC delta on average than those between calls.

\taggedpara{Online Phase} We run the enclave in debug mode to track true positives (interrupts when the nonce is biased) and false positives (interrupts when the nonce is not biased). We observe, on average, one interrupt per 300 to 1000 signatures, depending on the chosen IPI delays.

The ratio between true and false positives varies from one consecutive run to another. This is most likely due to system noise, combined with the small instruction delta. After filtering out unwanted \texttt{call} instructions, an average of 50\% to 60\% of flagged signatures were biased, depending on the current system state and chosen IPI delay. Two runs showed above-average true-positive rates, with $\geq75\%$ of flagged signatures being biased, but we could not consistently reproduce these results.

\taggedpara{Evaluation} The LBMS key leakage attack in~\ref{section:tdx} had few false positives and required fewer biased nonces, enabling fast subset-based lattice reduction attacks. Here, false positive rates of 40–50\% increase the cost for successful key recoveries. For a 34-all-biased subset from 500 signatures with a 50\% true positive rate, the expected number of reductions is $5.72 \times 10^{10}$, and $7.74 \times 10^{7}$ for a 60\% rate. At 60 BKZ reductions/sec on 16 cores, this corresponds to ~30.3 CPU years and 15 CPU days, respectively. Despite the higher cost compared to \ref{section:tdx}, the attack remains computationally feasible considering the potential for parallelization.

%% file: content/06-limitations.tex
\taggedpara{Limitations}
Probabilistic interrupt counting attacks have a smaller set of exploitable targets than deterministic single-stepping attacks. Our attacks require disabling caches and co-location of the attacker thread, which makes attacks that rely on single-stepping to amplify microarchitectural side-channels~\cite{DBLP:journals/iacr/HuangSCGJ23,frontal,platypus} potentially infeasible. Further, interrupt counting attacks typically rely on isolating vulnerable code patterns through state machines to track enclave progress across pages. Many such state machines do not require sampling the number of instructions between different states,  such as the one in~\ref{section:copycat} and the one used in the SECP160R1 attack~\cite{tdx-down}. However, some attacks require this additional granularity and construct augmented state machines that pair page accesses with per-page observed instruction counts, for example, an attack described in~\cite{DBLP:conf/uss/MoghimiBHPS20}. Neither PSS nor LBMS can provide single-trace accurate instruction counts, which reduces the potential targets to binaries for which secret-dependent control flow operations can be isolated using non-augmented state machines. 

PSS requires the secret to be static to average interrupt counts over multiple runs. While LBMS does not suffer from this limitation, it does require that the leaking branch be longer in terms of cycles than the alternate branch. Cycle requirements between secret-dependent branches are not relevant if the instruction delta is significant (Section~\ref{section:tdx}).

\taggedpara{Countermeasures} Interrupt-counting attacks (and related interrupt-driven attacks) could be mitigated by making SGX Enclaves uninterruptible. This, however, is currently not possible ~\cite{cui2023quanshield} and would require substantial architectural changes. Given that Intel SGX was not designed to process data from devices that may rely on interrupts, this approach could conceivably be implemented without fundamentally restricting Intel SGX's capabilities. Disabling interrupts may, however, have other consequences, such as enclaves hogging compute resources.

Another defense is implementing targeted mitigations against specific attack components. While this would prevent our current attacks, other techniques could potentially replace these building blocks and bypass such mitigations.

Our probabilistic interrupt counting attacks fundamentally depend on the attacker's ability to interrupt the enclave and precisely control the interrupt arrival time and distribution. Preventing an attacker from sharing a physical core with the enclave will make it harder for them to fingerprint the progress in the AEX-Notify mitigation, as shared resources and performance counters can no longer be used. Consequently, sending precisely timed IPIs will be harder. Intel advises to turn off hyperthreading in the BIOS of affected platforms in response to related vulnerabilities~\cite{intel_ht_advisory}. While this setting can be verified during the attestation process, it impacts the performance and utility of the system as it applies to the whole platform.  

Another avenue would be to harden an enclave's isolation. Currently, only thread-specific counters are disabled for SGX enclaves. This could be extended to all performance counters on a core. Additionally, hyperthreading could be disabled for physical cores hosting enclaves. Both approaches would make it harder for the attacker to fingerprint the mitigation and send precise interrupts. Another option would be to lock the enabled state for core-specific caches during enclave execution. This, however, does not exclude other methods from slowing enclave instructions.

Virtualization-based architectures~\cite{ahmad2023extensible} might provide additional tools to design interrupt-related mitigations, by mediating interrupts before delivering them to trusted code. However, as shown by attacks on architectures such as Intel TDX, these approaches also increase the complexity of the underlying TCB and thus offer an extended attack surface when considering interrupt-related attacks~\cite{tdx-down}.

%% file: content/07-related-work.tex
\taggedpara{Mitigating Interrupt-Driven Attacks} Several proposals have been made before the introduction of AEX-Notify to counter threats arising from interrupt-driven attacks by detecting suspicious interrupt rates~\cite{oleksenko2018varys, chen2017detecting, shih2017t}. However, these approaches suffer from inaccuracies due to the difficulty in determining a malicious interrupt frequency threshold for practical applications. Some approaches~\cite{chen2017detecting, shih2017t} furthermore rely on non-widely adopted hardware features (TSX). Defining malicious interrupt frequency heuristics~\cite{oleksenko2018varys} may prevent PSS attacks given the high rate of interrupts required. This does not apply to LBMS, which can be set up to fire at most one interrupt per trace. Mitigations aiming to prohibit interrupts by enforcing transactional memory operations using TSX~\cite{chen2017detecting, shih2017t} would counter many interrupt-related attacks, including ours.

Intel TDX includes a mitigation that aims to prevent single-stepping by activating a prevention mode if a trusted domain has not made enough progress between two interrupts. As shown in TDXDown~\cite{tdx-down}, this heuristic can be bypassed by tampering with the CPU's frequency. If the detection heuristic is triggered, TDX will execute a random number of instructions after registering an interrupt before returning to untrusted handlers. However, TDXDown uncovered a cache side-channel that deterministically reveals the number of instructions executed by the mitigation~\cite{tdx-down}. The TDXDown frequency tampering attack cannot be applied to AEX-Notify, as AEX-Notify does not use a heuristic-based approach to detect a lack of forward progress. Unlike TDXDown, our attacks do not require a side channel vector that leaks exact instruction counts between subsequent interrupts.

\taggedpara{Limiting Spatial Resolution} Interrupt-based attacks depend on coarse-grained 4 KiB side-channels to roughly determine the execution point of an enclave before initiating an attack. Such side-channels take the form of observing PTE page table attributes~\cite{DBLP:conf/uss/BulckWKPS17} or rely on page-fault sequences by selectively removing page permissions during an enclave's execution~\cite{xu2015controlled}. Attempts have been made to prevent these leakages~\cite{DBLP:conf/ccs/ShindeCNS16, vanoverloop2025tlblur}. While~\cite{DBLP:conf/ccs/ShindeCNS16} relies on extensive architectural changes to Intel SGX,  TLBlur~\cite{vanoverloop2025tlblur} works by extending AEX-Notify to prefetch the $N$ most recent application pages into the TLB with maximum permissions, effectively blurring the sequences of accessed pages during an enclave's lifetime. These mitigations reduce an attacker's spatial resolution, making it harder to pinpoint when to launch interrupt-based attacks.

\taggedpara{Performance Counters} Performance counters have been leveraged in prior work to assist with attacks on Intel SGX. Monitoring cache misses on a concurrent attacker thread~\cite{cache-attacks-are-practical} showed that attackers could indirectly fingerprint an enclave's memory patterns. Performance counters were shown to record events triggered during transient executions by instructions running in SGX enclaves by~\cite{qiu2022pmu}. Undocumented performance counters have been explored in an offensive setting by~\cite{yang2023exploration}. Similarly to~\cite{cache-attacks-are-practical}, our approach uses performance counters to monitor the concurrent attack thread, which shares resources with the enclave. Other approaches~\cite{qiu2022pmu, lee2024poster} mainly rely on repeatedly interrupting an enclave and sampling performance counters on the enclave thread, which is subject to restrictions imposed by Intel's ASCI mitigation, which prevents updates to thread-specific performance counters while an enclave is running.

%% file: content/08-conclusion.tex
AEX-Notify was designed to prevent Deterministic Single Stepping (DSS), and to the best of our knowledge, this security guarantee remains in effect. However, our work demonstrates that the defensive approach chosen for AEX-Notify has its limitations. AEX-Notify is based on two key ingredients. First, it reduces the execution window of the EARP by priming the microarchitectural state. Second, it builds the mitigation such that the adversary should not be able to learn whether the next interrupt lands during the mitigation or one of the subsequent enclave instructions. We have shown that both of these key ingredients can be manipulated. We extend the execution window of the enclave instructions simply by disabling caching. Even if our specific technique, cache disabling, were prevented, there are likely other ways to extend the execution window, given the complexity of the microarchitectural state and the numerous ways to prime it. We filter interrupts that occur during the mitigation, thereby breaking AEX-Notify's obfuscated forward progress guarantee, by observing the microarchitectural state changes caused by the mitigation execution from a spy thread. Again, even if our specific attack technique were prevented (e.g., by disallowing hyper-threading), it is likely that there are various other ways to observe the microarchitectural state changes caused by the mitigation's execution.

We conclude that both AEX-Notify's key ingredients are, and most likely will remain, vulnerable. In this paper, we have leveraged such manipulations to realize practical probabilistic interrupt counting attacks. It remains an open research question whether similar manipulation can be leveraged to implement DSS.